\documentclass[aps,prl,twocolumn, superscriptaddress]{revtex4-2}

\usepackage{times}
\usepackage{amsmath}
\usepackage{amssymb}
\usepackage{xfrac}
\usepackage{dsfont}
\usepackage{graphicx} 
\usepackage{float}
\usepackage{dcolumn} 
\usepackage{bm} 
\usepackage{braket}
\usepackage{mathtools}
\usepackage{xcolor}
\usepackage{hyperref}
\hypersetup{
     colorlinks=true,
     linkcolor=blue,
     filecolor=blue,
     citecolor=black,
     urlcolor=black,
     }
\usepackage[all]{hypcap}

\begin{document}

\title{Chaos-assisted Turbulence in Spinor Bose-Einstein Condensates}

\author{Jongmin Kim}
\thanks{These authors contributed equally to this work}
\affiliation{Department of Physics and Astronomy, Seoul National University, Seoul 08826, Korea}

\author{Jongheum Jung}
\thanks{These authors contributed equally to this work}
\affiliation{Department of Physics and Astronomy, Seoul National University, Seoul 08826, Korea}

\author{Junghoon Lee}
\affiliation{Department of Physics and Astronomy, Seoul National University, Seoul 08826, Korea}

\author{Deokhwa Hong}
\affiliation{Department of Physics and Astronomy, Seoul National University, Seoul 08826, Korea}

\author{Y. Shin}
\email{yishin@snu.ac.kr}
\affiliation{Department of Physics and Astronomy, Seoul National University, Seoul 08826, Korea}
\affiliation{Institute of Applied Physics, Seoul National University, Seoul 08826, Korea}


\begin{abstract}

We present a turbulence-sustaining mechanism in a spinor Bose-Einstein condensate, which is based on the chaotic nature of internal spin dynamics.
Magnetic driving induces a complete chaotic evolution of the local spin state, thereby continuously randomizing the spin texture of the condensate to maintain the turbulent state. 
We experimentally demonstrate the onset of turbulence in the driven condensate as the driving frequency changes and show that it is consistent with the regular-to-chaotic transition of the local spin dynamics.
This chaos-assisted turbulence establishes the spin-driven spinor condensate as an intriguing platform for exploring quantum chaos and related superfluid turbulence phenomena.

\end{abstract}

\maketitle

Turbulence, a ubiquitous phenomenon in fluids, poses a major challenge in physics owing to its complexity~\cite{feynman1971feynman}. To effectively investigate this phenomenon, it is essential to generate turbulence in a controlled manner. Various existing methods, such as moving grids~\cite{hideharu1991realization,mydlarski1996onset,smith1993decay}, impinging jets~\cite{cooper1993impinging,kim2023experimental}, and rotating drums~\cite{seiden2011complexity,eltsov2009turbulent,PhysRevLett.124.124501}, have provided valuable insights into the different aspects of turbulent states, highlighting the strong connection between the generation method and flow dynamics. However, these approaches are primarily based on external forces and inertial energy cascades~\cite{vinen2002quantum,navon2016emergence,alexakis2018cascades}, which might limit the possibilities of exploring novel and unconventional turbulent regimes. In this study, we propose an alternative approach to generate turbulence by harnessing the intrinsic chaos within the fluid itself. By utilizing the unpredictable and sensitive nature of chaotic systems, this method introduces energy directly into the flow through self-amplifying mechanisms such as the stretching and folding of fluid elements~\cite{reichl2004transition}. This chaos-based method holds the potential for improved mixing and the creation of a more energetic and sustained turbulent state, thereby offering the opportunity to explore previously uncharted turbulent regimes.

In this Letter, we present a case study of chaos-assisted turbulence in a driven spin-1 atomic Bose-Einstein condensate (BEC). Previous studies have observed a stationary turbulent state in the BEC when it is subjected to continuous radio frequency (rf) magnetic fields~\cite{PhysRevA.108.013318,PhysRevA.108.043309}. Under specific driving conditions, the turbulence attains its maximum intensity, accompanied by an isotropic spin composition. Although the oscillating magnetic field, aided by noise, has been identified as the driving force, the mechanism that sustains the generated turbulence while maintaining the isotropic spin nature has remained unclear. Here, through numerical simulations and experimental validation, we demonstrate that the applied magnetic field induces chaotic motion within local spin states, thus sustaining the turbulent state. 
This finding concretizes the notion of chaos-assisted turbulence generation, particularly in a superfluid system, paving the way for a deeper understanding and control of this complex and fascinating phenomenon.

A superfluid with internal spin degrees of freedom presents a departure from conventional superfluid dynamics. The order parameter of the superfluid can be expressed as $\Psi=\sqrt{n}e^{i\varphi}\boldsymbol{\zeta}$, where $n$ is the superfluid density, $\varphi$ is the superfluid phase, and $\boldsymbol{\zeta}$ is the spinor for the spin state~\cite{kawaguchi2012spinor}. 
Due to the intricate relation between the superfluid phase and the spin rotation, mass superflow is associated not only with the spatial variations of $\varphi$ but also with those of $\boldsymbol{\zeta}$, i.e., with the spin texture. The superfluid velocity and its vorticity are given by
\begin{eqnarray}
    \boldsymbol{v}_s= \frac{\hbar}{m}(\nabla \varphi - i\boldsymbol{\zeta}^{\dagger}\nabla\boldsymbol{\zeta}), \nonumber \\
    \nabla \times \boldsymbol{v}_s = -\frac{i\hbar}{m} \nabla \boldsymbol{\zeta}^{\dagger} \times \nabla \boldsymbol{\zeta},
    \label{superfluid}
\end{eqnarray}
where $m$ is the particle mass and $\hbar$ is the reduced Planck constant, $h/2\pi$.
Therefore, a mass superflow can be generated by manipulating the spin texture in space.

In this work, we consider a situation in which the local dynamics of $\boldsymbol{\zeta}$ is in a classically chaotic regime. Then, because of the sensitivity on initial conditions, small spatial fluctuations in a superfluid, even starting with a uniform spin texture, would develop into complex spatial variations of the spin states over time evolution, leading to an irregular spin texture. If the spin texture is kept irregular because of the chaotic nature of the local spin dynamics, the resulting turbulent flow in the superfluid would be sustained. This is a scenario of the aforementioned chaos-assisted turbulence generation.

\begin{figure}[t]
    \includegraphics[width=8.6cm]{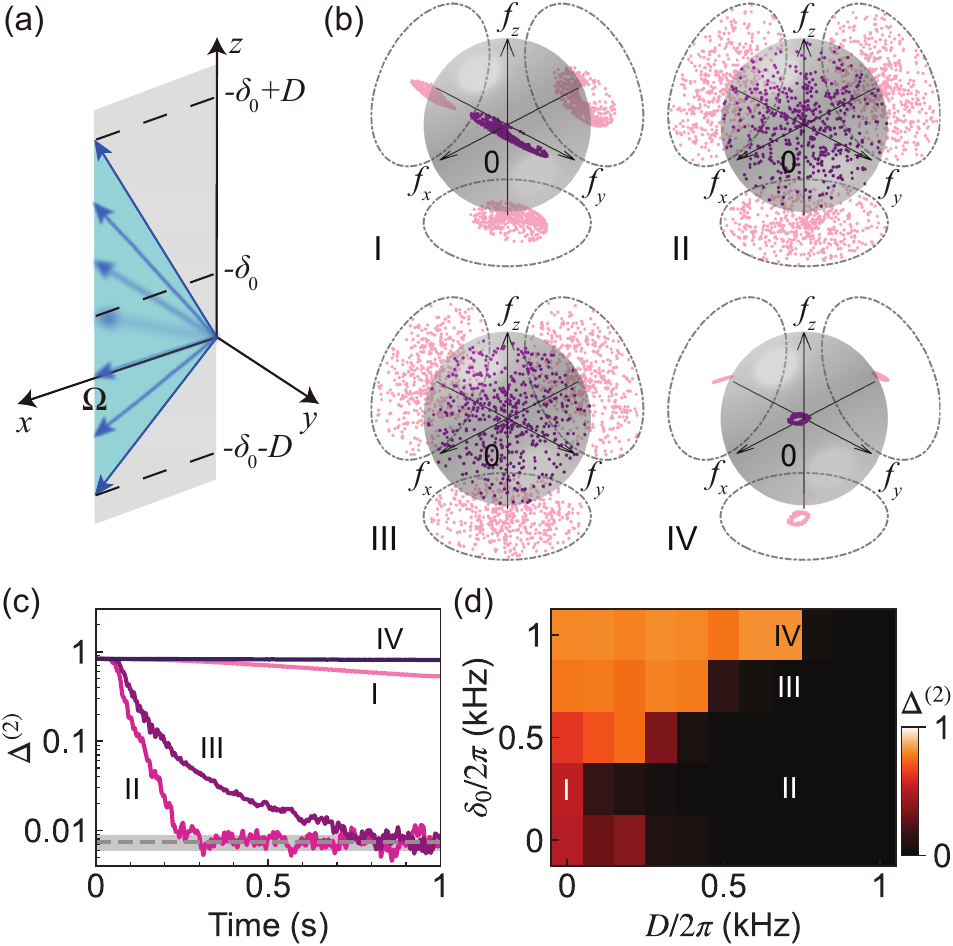}
    \centering
\caption{Chaotic spin dynamics in a spinor Bose-Einstein condensate under magnetic field modulations. (a) The linear Zeeman field in Eq.~(\ref{Hamiltonian}) oscillates as $\Omega \hat{\bf{x}}-\delta(t)\hat{\bf{z}}$ with $\delta(t)=\delta_0+D \sin (2\pi \nu t+\phi)$ and $\nu=60$~Hz. (b) Magnetization, $\boldsymbol{f}=(f_x,f_y,f_z)$, distributions of a spin state ensemble after 1-s spin driving for four different sets of driving parameters: (I) $\{\delta_0,D\}/2\pi = \{250,0\}$, (II) $\{250,700\}$, (III) $\{750,700\}$, and (IV) $\{1000,700\}$~Hz. The ensemble consists of $128^2$ spin states, which initially have $\boldsymbol{f}\approx 0$ (see text for details). (c) Time evolution of the second-momentum trace distances $\Delta^{(2)}$ between the spin state ensemble and the Haar random ensemble for (I)-(IV). The dashed line represents the expectation value for a set of $128^2$ spin states sampled from the Haar ensemble. (d) $\Delta^{(2)}$ after 1-s spin driving as a function of $\delta_0$ and $D$. 
}
    \label{FIG1}
\end{figure}

Our superfluid system is a BEC of $^{23}$Na atoms in the $F$=1 hyperfine state with antiferromagnetic interactions~\cite{zhou2003quantum,seo2015half}. 
A uniform external magnetic field of $B_z$ is applied along $\hat{\bm{z}}$, and for spin driving, an rf magnetic field with oscillating frequency of $\omega$ is applied along the transverse direction. 
In a mean-field description, neglecting the spatial modes of the BEC and taking the rotating wave approximation, the local dynamics of the spin state $\boldsymbol{\zeta}=(\zeta_{+1}, \zeta_{0},\zeta_{-1})^\text{T}$ is described by the following Hamiltonian per particle,
\begin{equation}
 H_s = \hbar \delta f_z  -\hbar \Omega f_x + q \boldsymbol{\zeta}^{\dagger}\text{f}_z^2\boldsymbol{\zeta} + \varepsilon_s \vert \boldsymbol{f} \vert^2,
 \label{Hamiltonian}
 \end{equation} 
where $\textbf{f} = (\text{f}_x,\text{f}_y,\text{f}_z)$ are the spin operators of the spin-1 system and $\boldsymbol{f} = \boldsymbol{\zeta}^{\dagger} \textbf{f} \boldsymbol{\zeta}$ is the spin vector with $f_{x,y,z}$ representing the magnetizations in $x,y,$ and $z$ directions, respectively. $\delta=\omega-\omega_0$ is the frequency detuning of the rf magnetic field from the Larmor frequency $\omega_0= \frac{1}{2} \mu_\text{B} B_z/\hbar$ with $\mu_\text{B}$ being the Bohr magneton, $\Omega$ is the Rabi frequency of the rf field, and $q$ denotes the quadratic Zeeman energy. The last term represents the energy of spin interaction, which introduces nonlinearity to the system.

When the energy scales of $\varepsilon_s$, $q$, and $\hbar\Omega$ are comparable and $\delta=0$, it is known that the spin dynamics of the Hamiltonian $H_s$ becomes chaotic~\cite{PhysRevA.101.053604,PhysRevLett.126.063401}. 
Here, we investigate a magnetic driving scheme where the external magnetic field $B_z$ is modulated such that $\delta(t)=\delta_0 + D\sin{(2\pi \nu t+\phi)}$ [Fig.~\ref{FIG1}(a)].
This field modulation breaks the energy conservation constraint, possibly enhancing the chaoticity of the system and facilitating complete randomization of the spin state~\cite{choi2023preparing,PRXQuantum.4.010311}.

To demonstrate the chaotic behavior of the spin system,
we numerically investigate the time evolution of an ensemble of spin states, $\mathcal{E}$, and characterize it in the coordinate space of magnetization $\boldsymbol{f}$.
The ensemble $\mathcal{E}$ consists of $128^2$ spin states, whose initial states are constructed from the $m_F=0$ state ($\boldsymbol{\zeta}=(0,1,0)^\text{T}$) with small Gaussian random noise added~\cite{SM}, so $\boldsymbol{f}\approx 0$.
The system parameters are set to $\varepsilon_s/h = 45$ Hz, $q/h=47$ Hz, $\Omega/2\pi = 200$ Hz and $\nu=60$~ Hz. Figure~\hyperref[FIG1]{1(b)} shows the $\boldsymbol{f}$ distributions of the ensemble after the evolution of 1 s for various sets of driving parameters $\{\delta_0,D\}$. Without field modulation ($D=0$) and for small $\delta_0$ (case I), the magnetization is dispersed on the plane perpendicular to the linear Zeeman field, but far from fully randomized. Meanwhile, when subjected to the field modulation, the ensemble spreads over a broader region of the magnetization space (cases II and III), implying the enhanced spin mixing of the system.~\cite{SM}

As a proxy of the randomness of the spin states in $\mathcal{E}$, we estimate a trace distance, 
\begin{eqnarray}
    \Delta^{(2)} \equiv \frac{1}{2}\vert\vert \rho_{\mathcal{E}}^{(2)}-\rho_{\text{Haar}}^{(2)} \vert\vert_{1},
    \label{QuantumStateDesign}
\end{eqnarray}
where $\rho_{\mathcal{E}}^{(2)}=\sum_{\boldsymbol{\zeta} \sim \mathcal{E}} \big( \boldsymbol{\zeta} \boldsymbol{\zeta}^{\dagger} \big)^{\otimes 2}$ is the second momentum of the ensemble and $\rho_{\text{Haar}}^{(2)}$ is that of the Haar random ensemble $\mathcal{E}_\text{Haar}$, which is a unitarily invariant, maximally randomized ensemble of the spin-1 system~\cite{PhysRevLett.128.060601, PRXQuantum.4.010311, PRXQuantum.4.030322}. 
Here, $\vert\vert \cdot \vert\vert_{1}$ denotes the trace norm, and $\Delta^{(2)}$ represents the difference from the fully random ensemble in second momentum. 
Figure~\hyperref[FIG1]{1(c)} shows the evolution curves of $\Delta^{(2)}$ for the different driving parameters.
In a specific driving condition, $\mathcal{E}$ becomes fully randomized as $\Delta^{(2)} \simeq 0$.
In Fig.~\hyperref[FIG1]{1(d)}, the value of $\Delta^{(2)}$ after 1~s of spin driving is displayed in the plane of $\delta_0$ and $D$.
Interestingly, the dynamic behavior of the system exhibits a sudden transition from chaotic to regular as $\delta_0$ exceeds a threshold value $\delta_\text{th}$.
The value of $\delta_\text{th}$ is found to be linearly proportional to the magnitude of field modulation as $\delta_\text{th} \simeq D$, highlighting the role of field modulation in the randomization of the spin state.

\begin{figure}[t]
    \includegraphics[width=8.6cm]{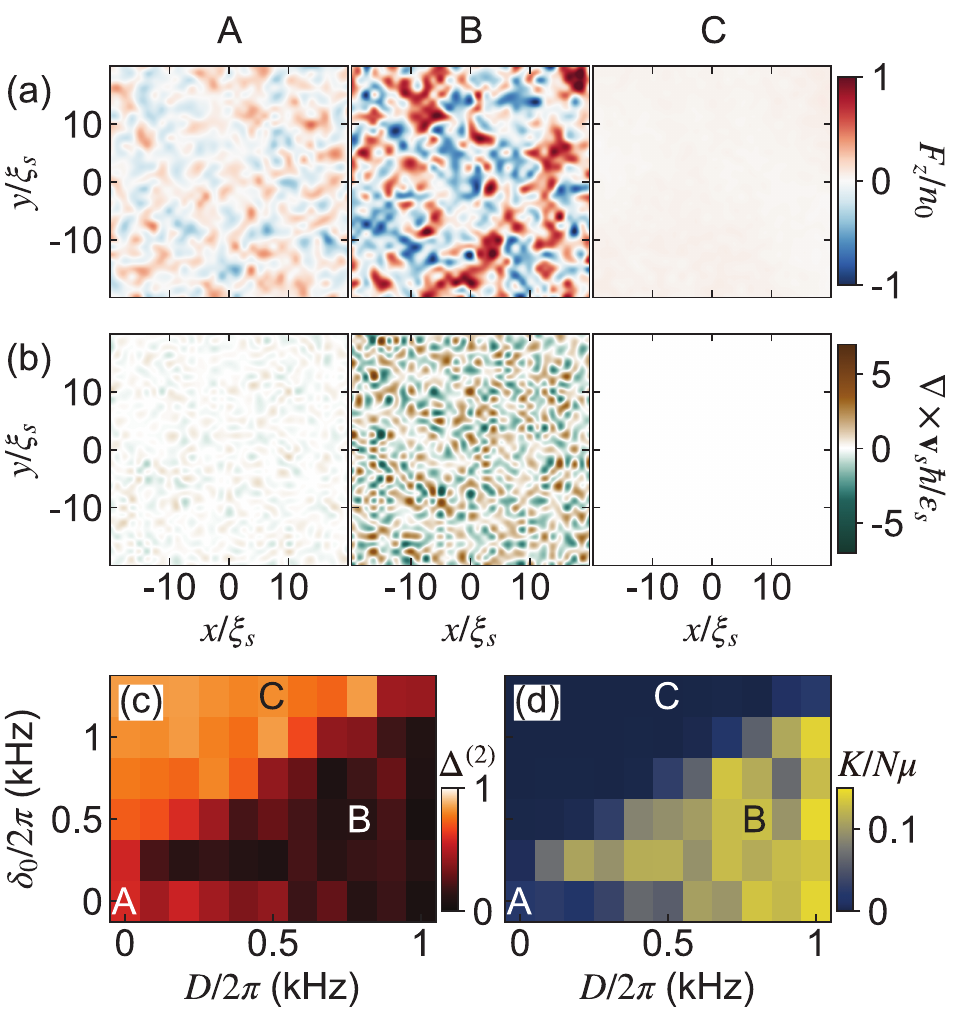}
    \centering
\caption{Turbulence generation in a spinor BEC by spin driving. (a) Magnetization $F_z$ and (b) vorticity $\nabla\times \boldsymbol{v}_s$ distributions in a two-dimensional BEC after 1-s spin driving for three different driving conditions: (A) $\{\delta_0, D\}/2\pi=\{0,0\}$, (B) $\{500,800\}$, and (C) $\{1250, 500\}$~Hz. $F_z=nf_z$ is the density of magnetization along the $z$ direction and $\boldsymbol{v}_s$ is the superflow velocity. $n_0$ is the mean particle density, $\varepsilon_s$ denotes the spin interaction energy scale, and $\xi_s$ is the spin healing length.
(c) $\Delta^{(2)}$ and (d) kinetic energy $K$, as functions of $\delta_0$ and $D$. $N$ is the total particle number and $\mu$ denotes the chemical potential of the BEC.}
    \label{FIG2}
\end{figure}

We extend our discussion to the dynamics of a BEC with spatial extent, e.g., in two dimensions. 
Here, the spinor condensate can be viewed as the spatial array of the local nonlinear spin subsystems that are coupled to their neighbors. 
We numerically calculate the system's evolution using spin-1 Gross-Pitaevskii equations (GPEs)~\cite{SM,footnote1_gridsize}.
We prepare the initial condensate in the $m_F=0$ state with small random noises~\cite{PhysRevA.108.043309,SM}, and observe that an irregular spin texture develops under spin driving, e.g., with $\delta_0/2\pi=500$ Hz and $D/2\pi=800$ Hz [Fig.~\hyperref[FIG2]{2(a)} case B]. 
The irregular spin texture implies turbulent superflow in the spinor BEC, as verified by the vorticity distribution of the superfluid velocity in Fig.~\hyperref[FIG2]{2(b)}.

The range of driving parameters for the generation of turbulence is well aligned with that for the chaotic spin dynamics. 
In Figs.~\hyperref[FIG2]{2(c)} and \hyperref[FIG2]{2(d)}, we plot the trace distance $\Delta^{(2)}$ and the kinetic energy $K$ of the system in the $\delta_0$-$D$ plane, respectively. Here, $\Delta^{(2)}$ is measured from the spin state ensemble $\mathcal{E}$ obtained by position projection of the condensate~\cite{PhysRevA.108.043309} and the kinetic energy is calculated as $K = \int d^2 \boldsymbol{r} \Psi^{\dagger} (-\frac{\hbar^2}{2m}\nabla^2)\Psi$. 
The relationship of $\delta_\text{th} \simeq D$ is still observed as in Fig.~\hyperref[FIG1]{1(d)}, which clearly demonstrates that the turbulence generation originates from chaotic spin dynamics. 
Meanwhile, the value of $\Delta^{(2)}$ for turbulent BEC appears to be relatively higher than the corresponding value in Fig.~\hyperref[FIG1]{1(d)}. It is attributed to the energy dissipation process through spin-wave relaxation~\cite{PhysRevA.108.013318}, which would homogenize the spin texture. 
Further characterization of the velocity fields of the turbulent BEC is provided in~\cite{SM}, where a $-5/3$ power-law scaling was observed in the incompressible part of the kinetic energy~\cite{fujimoto2014spin}.


\begin{figure}[t]
    \includegraphics[width=8.6cm]{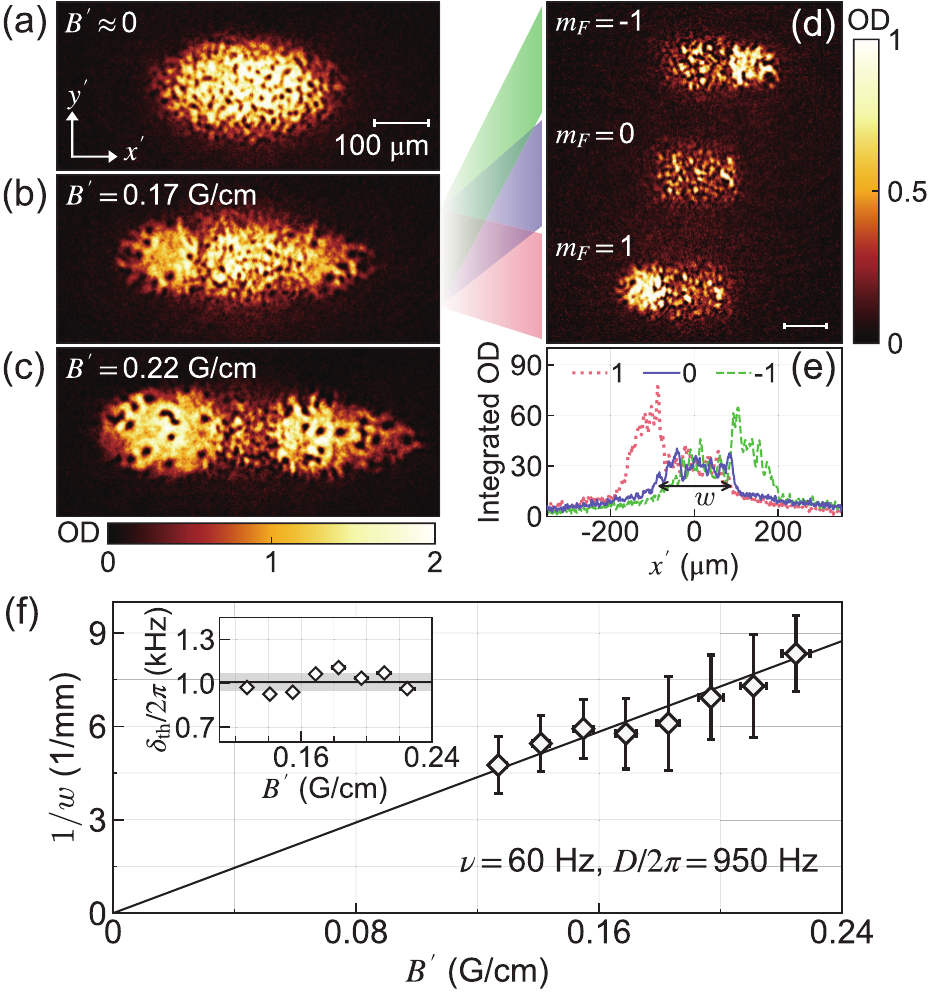}
    \centering
\caption{Observation of the turbulence onset in a driven spinor BEC. (a)-(c) Time-of-flight images of BECs after spin driving for 2 s with $\delta_0=0$, $D/2\pi=950$~Hz, $\nu=60$~Hz, and $\phi=0$. Turbulence appears as irregular density patterns in the freely expanding BECs. A magnetic field gradient $B'$ was applied along the $x'$ direction during spin driving. (d) Image after Stern-Gerlach spin separation for $B'=0.17$~G/cm. The deformation of the clouds due to separation is compensated by perspective transformation~\cite{SM}. (e) Integrated density profiles of the three spin components. The width $w$ of the spin turbulence region was determined from the density distribution of the $m_F=0$ component. (f) $1/w$ as a function of $B'$. Each data point was obtained from the twenty to thirty measurements and its error bar indicates their standard deviation. The solid line is a linear line of $w^{-1}=\alpha B'$ with $\alpha=3.7\times 10^4$/G, fit to the data. The inset shows the values of $\delta_\text{th}=\frac{\mu_B}{4\hbar}B'w$. The  horizontal line and the shaded region indicate their mean and standard deviation, respectively.}
    \label{FIG3}
\end{figure}

Next, we conduct an experimental verification of the turbulence generation mechanism assisted by chaotic spin dynamics. We prepare a BEC consisting of about $3.5\times$10$^{6}$ $^{23}$Na atoms in the $|F$$=$$1, m_{F}$$=$$0\rangle$ state. The BEC is trapped in an optical dipole trap of highly oblate geometry and its Thomas-Fermi (TF) radii are $(R_{x'},R_{y'},R_{z'})$~$\approx$~$(160, 75, 1.6)~\mu$m, where $x', y'$ and $z'$ denote spatial coordinates. For the peak atomic density at the trap center, the spin interaction energy is $\varepsilon_s \approx h \times 53$~Hz~\cite{PhysRevA.83.042704} and the spin healing length $\xi_s=\hbar/\sqrt{2m\varepsilon_s}\approx2.0~\mu$m, comparable to the thickness of the sample. A uniform external magnetic field of $B_z\approx 0.41$~G is applied along $\hat{\bm{z}}=(-\hat{\bm{x}}'+\hat{\bm{y}}')/\sqrt{2}$, giving $\omega_0\approx 2\pi\times 291$~kHz and $q \approx h\times 47$~Hz. 
An rf magnetic field is applied along $\hat{\bm{y}}'$ with $\Omega=2\pi \times 150$~Hz. The external magnetic field $B_z$ is sinusoidally modulated at $\nu=60$~Hz with a variable magnitude of less than a few mG.

Figure~\hyperref[FIG3]{3(a)} shows an optical density (OD) image of a BEC, taken along the $\hat{\bm{z}}'$ direction after a 18-ms time-of-flight~\cite{footnote3_SGimage}. The BEC was spin-driven for 2~s with $\delta_0=0$ and $D/2\pi=950$~Hz. A stationary turbulent state appears in the driven BEC, manifested by an irregular density distribution that arises after free expansion from the chaotic velocity field of the driven condensate. Furthermore, the turbulent BEC exhibits equal populations in all three spin states. The spin isotropy of the turbulent BEC was demonstrated in Ref.~\cite{PhysRevA.108.013318} by showing that the populations of the three spin states are equal regardless of the quantization axis.

We investigate how the frequency detuning $\delta_0$ of the rf driving field affects turbulence generation by conducting a similar experiment with an additional magnetic field gradient $B'$ applied along the long axis of the condensate. 
This field gradient renders the frequency detuning of the rf driving spatially varing across the condensate as $\delta_0(x')=(\frac{1}{2}\mu_\text{B} B' x')/\hbar$. 
We observe that spin-isotropic turbulence occurs within a central region of the condensate [Fig.~\hyperref[FIG3]{3(b)} and \hyperref[FIG3]{3(c)}], whose spatial extent decreases as $B'$ increases, implying the existence of a range of $\delta_0$ effective for the generation of turbulence.

The image obtained after Stern-Gerlach spin separation reveals that the turbulent region is positioned between two ferromagnetic domains with opposite magnetizations [Fig.~\hyperref[FIG3]{3(d)}]. 
For high $|\delta|$, the ground state of the driven BEC is ferromagnetic owing to the large Zeeman energy. As the BEC is driven, the ferromagnetic domains gradually appear at both ends of the condensate and reach saturation in the steady state, containing vortices~\cite{SM}. 
The interface between the turbulent region and the ferromagnetic domains is clearly defined, showing sudden changes in density for the different spin components [Fig.~\hyperref[FIG3]{3(d)} and \hyperref[FIG3]{3(e)}]. This allows us to reliably determine the turbulent region from the density distribution of the $m_F=0$ component.

We measure the spatial extent $w$ of the spin-isotropic turbulent region along the direction of the field gradient, and find that its inverse is proportional to the magnitude of the field gradient [Fig.~\hyperref[FIG3]{3(f)}].
This means that the threshold value of $\delta_0$ for turbulence generation is consistently determined as $\delta_\text{th}=\delta_0(\frac{w}{2})=\mu_\text{B} B'w /4\hbar$. 
The presence of such a threshold detuning is in excellent agreement with the numerical results shown in Fig.~\hyperref[FIG2]{2}, providing clear evidence of the regular-to-chaotic transition in the driven spinor BEC system.

\begin{figure}[t]
    \includegraphics[width=8.6cm]{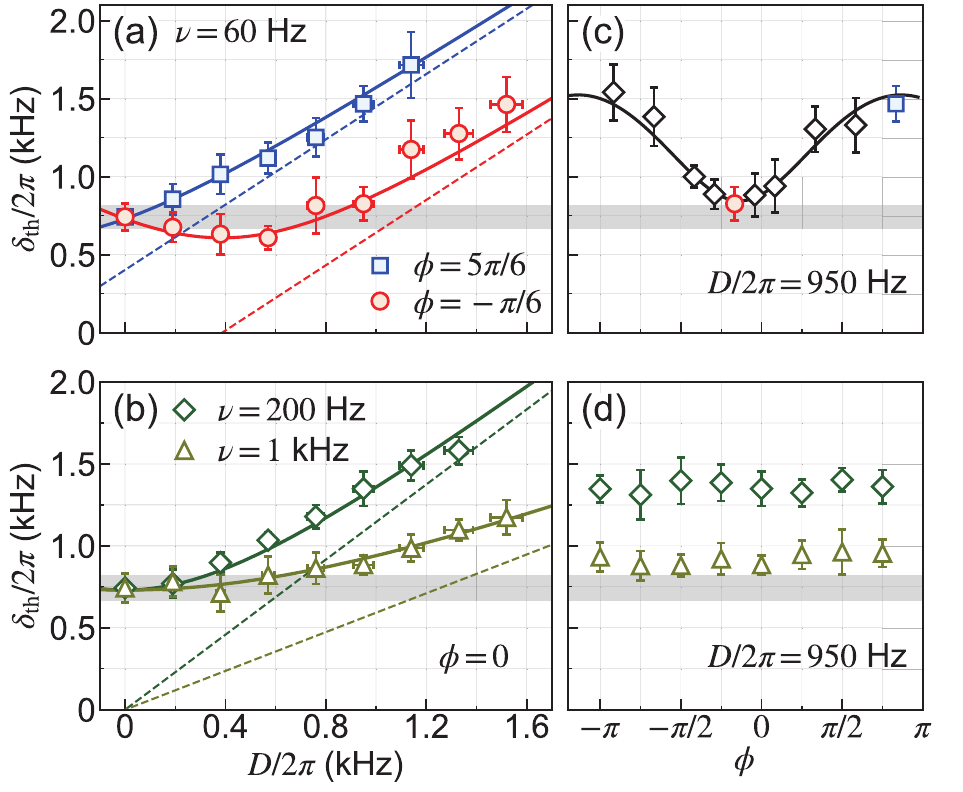}
    \centering
\caption{Threshold detuning $\delta_\text{th}$ of the spin driving for turbulence generation. Dependence of $\delta_\text{th}$ on the driving amplitude $D$ for (a) $\nu=60$~Hz , (b) 200~Hz and 1~kHz. The gray horizontal lines indicate the threshold detuning value of $\delta_{\text{th},0}/2\pi=0.74(9)$~kHz due to ambient field noise. Corresponding $\delta_\text{th}$ data for different modulation phases $\phi$ with $D/2\pi=1$~kHz for (c) $\nu=60$~Hz , (d) 200~Hz and 1~kHz. 
Each data point indicates the mean value of ten to twenty measurements, and its error bar is the standard deviation of the measurements.
The solid lines in (a) and (c) are the model curves of Eq.~(\ref{model}) with the parameter values of $\{\beta_\nu, D_\nu/2\pi, \phi_\nu\}=\{1.05, 0.39~\text{kHz}, 0.86\pi\}$, which were determined from a model fit to the data in (c). The solid lines in (b) are model curves fit to the data with $D_\nu=0$. The dashed lines indicate the corresponding model curves with $\delta_{\text{th},0}=0$.
} 
    \label{FIG4}
\end{figure}

The relationship of $\delta_\text{th}$ and the magnitude $D$ of the field modulations is investigated (Fig.~\hyperref[FIG4]{4}). As $D$ increases, in general, $\delta_\text{th}$ increases, which is consistent with the expected behavior of the spin dynamics. However, we observe that a turbulent region is formed even in the absence of field modulations, indicating a threshold detuning of $\delta_{\text{th},0}=0.74(9)$~kHz for $D=0$. We find that it is caused by ambient magnetic field fluctuations, which amounts to about 1~mG~\cite{PhysRevA.108.013318}.
Magnetic field noises are likely generated by current ripples in the magnetic coils that leak from the 60-Hz AC power line. In fact, for $\nu=60$~ Hz, $\delta_\text{th}$ shows a sinusoidal dependence on the phase $\phi$ of the external field modulation~[Fig.~\hyperlink{FIG4}{4(c)}], and it arises from interference with the background field component oscillating at the same frequency.  

Taking into account such ambient field fluctuation effects, we present a model of the threshold detuning as
\begin{equation}
\delta^2_\text{th}=\beta_\nu^2\big(D^2 + 2D D_\nu \cos(\phi-\phi_\nu)\big)+\delta_{\text{th},0}^2,
\label{model}
\end{equation}
where $D_\nu$ and $\phi_\nu$ represent the amplitude and phase of the background field that oscillates at frequency $\nu$, respectively, and $\beta_\nu$ denotes the proportionality of $\delta_\text{th}$ to the field modulation~\cite{SM}. Our measurement results of $\delta_\text{th}$ for various $D$ and $\phi$ are well described by the model [Fig.~\hyperref[FIG4]{4(a)}], suggesting $D_{\nu}/2\pi=0.39(5)$~kHz and $\beta_\nu=1.05(6)$ for $\nu=60$~Hz. 
In Fig.~\hyperref[FIG4]{4(b)}, we show the measurement data of $\delta_\text{th}$ for $\nu=200$~Hz and 1~kHz, where the model curve fitting to the data provides $D_\nu\approx 0$ and $\beta_\nu=1.15(4)$ and 0.59(4), respectively.  It is evident that when the modulation frequency surpasses the energy scales relevant to the system, the efficiency of the mixing caused by the external field modulation is reduced.

In conclusion, we have presented a turbulence generation mechanism based on chaos and demonstrated it with the spinor BEC system under magnetic field driving. The presence of the threshold frequency detuning $\delta_\text{th}$ and its linear relationship with $D$ were demonstrated, in excellent agreement with the predicted regular-to-chaotic transition. The stationary turbulence state of the spinor BEC provides interesting opportunities to explore nonequilibrium superfluid dynamics. An important next step is to compare this chaos-assisted turbulence with conventional hydrodynamic turbulence, where energy is injected on a certain length scale and an inertial energy cascade follows~\cite{vinen2002quantum,navon2016emergence,alexakis2018cascades,SM}. Since our driven system exhibits spatio-temporally chaotic flow induced by its internal spin dynamics, it might be discussed in the context of {\it active} turbulence~\cite{thampi2016active, alert2022active}, which is self-driven turbulence by spontaneous flow instability.
Further research should focus on quantifying the spin texture~\cite{fujimoto2012counterflow,fujimoto2013spin} and the mechanisms of energy transfer and dissipation~\cite{bradley2011direct,fujimoto2014spin} of the turbulent BEC.

\begin{acknowledgments}
This work was supported by the National Research
Foundation of Korea (Grants No. NRF-2023R1A2C3006565 and No. NRF-2023M3K5A1094811).

\end{acknowledgments}

\onecolumngrid
\pagebreak

\begin{center}
\large \textbf{Supplemental Material:\\
Chaos-assisted Turbulence in Spinor Bose-Einstein Condensates}\\
\end{center}

\setcounter{equation}{0}
\setcounter{figure}{0}
\setcounter{table}{0}
\makeatletter
\renewcommand{\theequation}{S\arabic{equation}}
\renewcommand*{\theHequation}{\theequation}
\renewcommand{\thefigure}{S\arabic{figure}}
\renewcommand*{\theHfigure}{\thefigure}
\renewcommand*{\bibnumfmt}[1]{[S#1]}

\subsection{A.\quad Numerical simulation of the spin dynamics}

From the Hamiltonian of Eq.~(\ref{Hamiltonian}), the equation of motion for the spin state of the BEC is given by 
\begin{eqnarray}
    i\hbar\partial_{t} \boldsymbol{\zeta} = \Big[ \hbar \delta(t) \text{f}_z +q \text{f}_z^2 - \hbar \Omega \text{f}_x + \varepsilon_s \sum_{i=x,y,z}f_i\text{f}_i \Big] \boldsymbol{\zeta}.
    \label{SMA_supp}
\end{eqnarray}
We numerically solved it using
a relaxation scheme \cite{Antoine15}, with $q/h=47$~Hz, $\Omega/2\pi=200$~Hz, $\varepsilon_s/h=45$~Hz for various driving conditions of $\delta(t)=\delta_0+D\sin(2\pi\nu t+\phi)$. The initial state was the easy-axis polar (EAP) state along the $z$ axis, that is, $\boldsymbol{\zeta}_0 = (0,1,0)^\text{T}$. To consider the effect of quantum noise, we added a Gaussian noise of $\kappa \boldsymbol{\eta}=\kappa (\eta_{1},\eta_{2},\eta_{3})^\text{T}$ to the initial state, where $\eta_{j} = \eta_{j,\text{R}}+i\eta_{j,\text{I}}$ ($j=1,0,-1$) are complex random variables with $\eta_{j,\text{R}}$ and $\eta_{j,\text{I}}$ being real random numbers of the standard normal distribution. We set $\kappa=0.005$~\cite{PhysRevA.108.043309}, which represents the ratio of the amount of noise to the particle density. In the analysis of the time evolution of a spin state ensemble, we consider a set of $128^2$ spin states whose noise components are chosen independently.

\begin{figure}[h]
    \includegraphics[width=17cm]{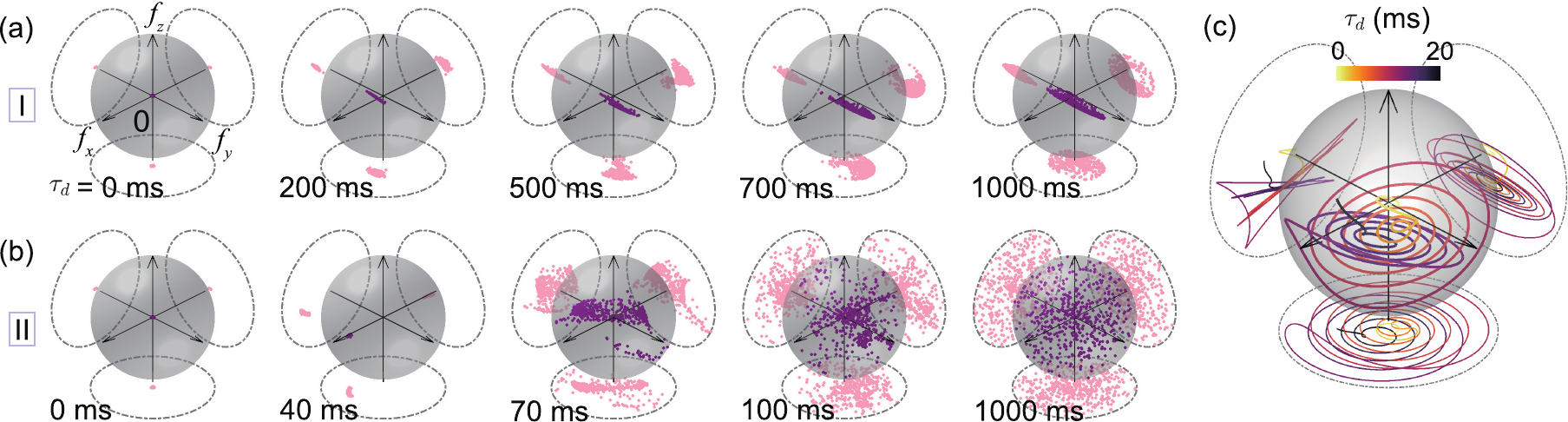}
    \centering
\caption{Magnetization $\boldsymbol{f}=\{f_x,f_y,f_z\}$ distributions of the spin state ensemble at various rf driving times $\tau_d$ for (a) $\{\delta_0, D\}/2\pi = \{250, 0\}$~Hz (case I in Fig.~1) and (b) $\{250, 700\}$~Hz (case II in Fig.~1), and $\nu=60$~Hz. (c) Trajectory of a spin state in the magnetization space for $0<\tau_d<20$~ms, where the initial state is $\boldsymbol{\zeta}_0=(0,1,0)^{\text{T}}$ and the driving parameters are the same in (b). The color of the trajectory denotes the evolution time $\tau_d$.}
    \label{FIG.S1}
\end{figure}

\subsection{B.\quad Gross-Pitaevskii equations of spin-1 Bose-Einstein condensates}

The dynamics of a two-dimensional BEC under spin driving was numerically investigated using spin-1 Gross-Pitaevskii equations (GPEs),
\begin{eqnarray}
    i \hbar \partial_{t} \Psi = \Big[ -\frac{\hbar^2}{2m}\nabla^{2}  + \hbar \delta(t) \text{f}_z + q \text{f}_z^2 - \hbar \Omega \text{f}_x 
    + g_n n + g_s\sum_{i=x,y,z}F_i \text{f}_i -\mu \Big] \Psi,  
    \label{GPE_supp}
\end{eqnarray}
where $g_n$ and $g_s$ denote the particle and spin interaction coupling constants, respectively, $\mu=g_n n_0=\gamma^2\varepsilon_s$ is the chemical potential of the condensate, and $\boldsymbol{F} = (F_x,F_y,F_z)=n(f_x,f_y,f_z)$ is the magnetization density. For $F=1$ $^{23}$Na BECs, \linebreak 
\vspace{10pt}
\rule{36pt}{0.05pt}
\vspace{1pt}\\
\noindent
{\footnotesize
$^{*}$ These authors contributed equally to this work\\
$^{\dagger}$ yishin@snu.ac.kr}

\pagebreak
\noindent
$\gamma = \sqrt{g_n/g_s}=5.3$. The initial state was the EAP state, $\Psi_0 = \sqrt{n_0}(0,1,0)^{\text{T}}$ with $n_0 = 10^4/\xi_s^2$ and noise was added. To incorporate quantum fluctuations, we used the truncated Wigner approximation~\cite{polkovnikov2010phase}. The noise distribution corresponds to the summation of a half quantum for each Bogoliubov excitation mode in the EAP state for $q\to\infty$~\cite{blakie2008dynamics}. The GPEs in Eq.~(\ref{GPE_supp}) were solved numerically using a relaxation pseudospectral scheme~\cite{Antoine15}. In Fig.~2, the size of the system was $160 \xi_s \times 160 \xi_s$, covered by a $256 \times 256$ grid of equally spaced points, where $\xi_s~\simeq~2.3$ $\mu$m.

\subsection{C.\quad Power spectrum of incompressible kinetic energy}

Tsubota {\it et al.}~\cite{fujimoto2012counterflow,fujimoto2013spin,fujimoto2014spin} investigated spin-superflow turbulence in ferromagnetic spin-1 BECs using analytical and numerical methods, and showed that the turbulent states arising from counterflow instability exhibit $-7/3$ and $-5/3$ power-laws in the spectra of the spin interaction energy and the incompressible part of the kinetic energy, respectively. In our numerical study of BEC dynamics under spin driving, we have observed that the same power-law behavior appears in the driven BEC with antiferromagnetic interactions, not only in the spin interaction energy, as described in our earlier work~\cite{PhysRevA.108.043309}, but also in the incompressible part of the kinetic energy.

For the analysis of the kinetic energy spectrum, we followed the method described in \cite{fujimoto2014spin}. The energy of superfluid velocity is given by $E_{v} = \frac{m}{2} \int d^2 \boldsymbol{r} \vert \boldsymbol{A}_v \vert^2$, where $\boldsymbol{A}_v = \sqrt{n}\boldsymbol{v}_s$. According to the Helmholtz theorem, the weighted velocity field $\boldsymbol{A}_v$ can be decomposed as $\boldsymbol{A}_v = \boldsymbol{A}_{iv}+\boldsymbol{A}_{cv}$ such that $\nabla \cdot \boldsymbol{A}_{iv} = 0$ and $\nabla \times \boldsymbol{A}_{cv} = \boldsymbol{0}$. Then, the kinetic energy is expressed as the sum of two parts as $E_{v} = E_{iv}+E_{cv}$ with $E_{\alpha v} = \frac{m}{2} \int d^2 \boldsymbol{r} \vert \boldsymbol{A}_{\alpha v} \vert^2$  ($\alpha \in \{i,c\}$). $E_{iv}$ ($E_{cv}$) is called the incompressible (compressible) part of the superfluid velocity energy. The power spectrum of $E_{iv}$ is obtained as follows:
\begin{eqnarray}
    \boldsymbol{A}_{iv}(\boldsymbol{r})&=&\sum_{\boldsymbol{k}} \tilde{\boldsymbol{A}}_{iv}(\boldsymbol{k}) e^{i\boldsymbol{k}\cdot\boldsymbol{r}} \nonumber \\
    E_{iv}(k) &=& \frac{Nm}{2n_0} \sum_{k<\vert \boldsymbol{k} \vert<k+\Delta k} \vert \tilde{\boldsymbol{A}}_{iv}(\boldsymbol{k}) \vert^2.
    \label{Eiv}
\end{eqnarray}

\noindent
Similarly, the power spectrum of the spin interaction energy $E_s = \frac{g_s}{2} \int d^2 \boldsymbol{r} \vert \boldsymbol{F}(\boldsymbol{r}) \vert^2$ is obtained as follows:
\begin{eqnarray}
    \boldsymbol{F}(\boldsymbol{r})&=& \sum_{\boldsymbol{k}} \tilde{\boldsymbol{F}}(\boldsymbol{k}) e^{i\boldsymbol{k}\cdot\boldsymbol{r}} \nonumber \\
    E_s(k) &=& \frac{N\varepsilon_s}{2n_0^2} \sum_{k<\vert \boldsymbol{k} \vert <k+\Delta k} \vert \tilde{F} (\boldsymbol{k}) \vert^2
    \label{Es}
\end{eqnarray}

\begin{figure}[h]
    \includegraphics[width=13.5cm]{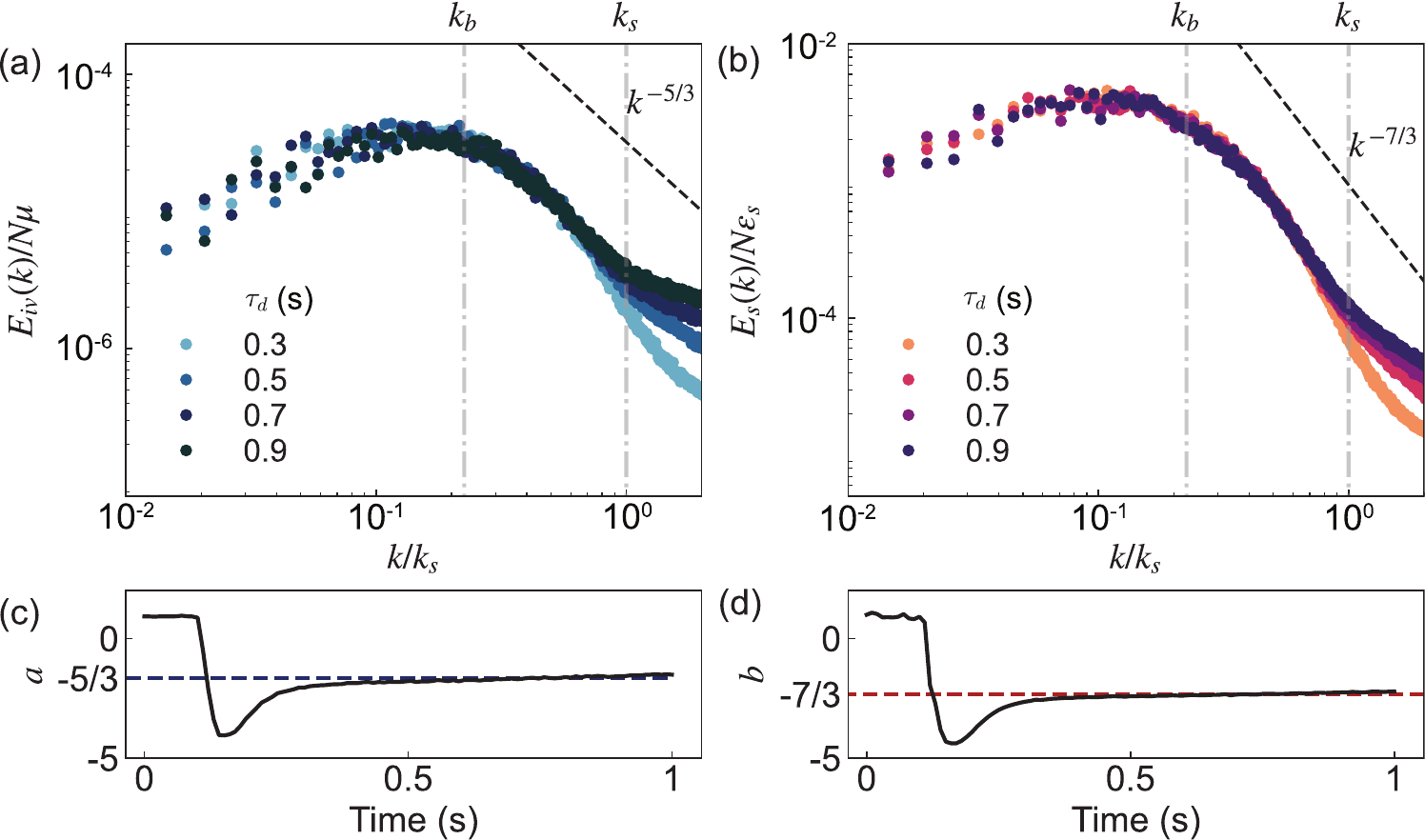}
    \centering
\caption{Energy spectrum and scaling behavior. Power spectra of (a) the incompressible kinetic energy $E_{iv}(k)$ and (b) the spin interaction energy $E_s(k)$ of a spin-driven BEC for various driving times. The $x$ and $y$ axes are on logarithmic scale. The black dashed lines represent $k^{-5/3}$ in (a) and $k^{-7/3}$ in (b), respectively. The power-law exponents were determined from the fitting of $E_{iv(s)}(k)=E' k ^{a(b)}$ to the data within the range of $k_b=\sqrt{2}/\xi_s<k<k_s=2\pi/\xi_s$, indicated by the vertical gray dashed-dotted lines. In (c) and (d), the resulting exponent values of $a$ and $b$ are displayed as functions of the driving time, respectively. For (b,d), we extracted the data from the study reported in Ref.~\cite{PhysRevA.108.043309}.}
    \label{FIG.S2}
\end{figure}

In Fig.~\ref{FIG.S2}, we present the energy spectra $E_{iv}(k)$ and $E_s(k)$ of the spin-driven BEC for different driving times, obtained from our numerical results using Eq.~(\ref{GPE_supp}) for $\{\delta_0,D\}/2\pi=\{0,1\}$~kHz and $\nu=60$~Hz. In the calculation, we used a grid of $1024 \times 1024$ configuration to expand the range of the wave number $k$~\cite{PhysRevA.108.043309}. When the driven BEC reaches a stationary turbulent state for $\tau_d>0.3$~s, $E_{iv}(k)$ and $E_s(k)$ approximately follow power laws with exponents of $-5/3$ and $-7/3$, respectively, within the range of $k_b<k<k_s$, where $k_s = 2\pi/\xi_s$ and $k_b = k_s/\sqrt{2}\pi$~\cite{fujimoto2013spin,PhysRevA.108.043309}. It is important to note that the observation of the $-5/3$ power-law scaling does not necessarily imply a direct energy cascade~\cite{fujimoto2014spin}.

\subsection{D.\quad Magnetic domain formation in a driven BEC}

\begin{figure*}[h]
    \includegraphics[width=18cm]{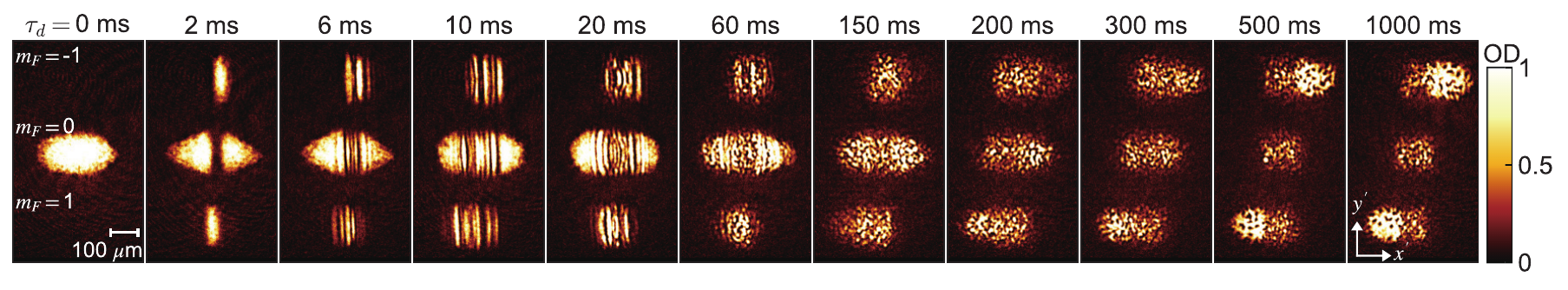}
    \centering
\caption{Turbulence generation and magnetic domain formation in a spin-driven BEC. Time-of-flight images of BECs after Stern-Gerlach spin separation for various rf driving times, $\tau_d$. A magnetic field gradient of $B'=0.14$~G/cm was applied along $\hat{\bm{x}}'$ before rf field driving.}
    \label{FIG.S3}
\end{figure*}

In Fig.~\ref{FIG.S3}, we present experimental images of the BECs for different spin-driving times, $\tau_d$, to demonstrate the process of magnetic domain formation under a field gradient. The initial BEC was prepared in the $m_F =0$ state, consisting of approximately $2.5\times10^{6}$ $^{23}$Na atoms. We applied a magnetic field gradient of $B'=0.14$ G/cm along the $x'$ direction and a rf driving field with $\Omega=2\pi\times150$ Hz. For short spin-driving times, $\tau_d\leq$10 ms, a stripe density pattern emerges in the central region of the BEC, indicating the development of a helical spin structure owing to spatial variations in frequency detuning as $\delta_0(x')=(\frac{1}{2}\mu_B B'x')/\hbar$. Subsequently, for 10 ms $<\tau_d<$200 ms, the stripe pattern breaks into smaller segments and the spin texture becomes irregular, starting from the center region and then spreading throughout the BEC. Meanwhile, the $m_F =\pm1$ spin components move in opposite directions owing to the field gradient $B'$, resulting in the buildup of net magnetization at both ends. After $\tau_d>200$ ms, the ferromagnetic domains, although still turbulent, become more distinct, and their boundary with the spin-isotropic turbulent region becomes more pronounced. Remarkably, the boundary exhibits a sudden change in the density of the corresponding spin component [Fig.~\ref{FIG3}(e)]. For $\tau_d>0.5$~s, the driven BEC reaches a state with a stationary magnetic domain structure, where the spin-turbulent region is sandwiched between two ferromagnetic domains.

\subsection{E.\quad Image reconstruction for compensating Stern-Gerlach spin separation}

In the Stern-Gerlach (SG) spin separation imaging method, we applied a magnetic field gradient pulse right after releasing the trapping potential and before imaging. This procedure enabled the spatial separation of spin components during the time of flight, allowing for examination of the spin texture of the BEC. However, the non-uniformity of the field gradient hinders a precise quantitative analysis of the spin texture because it causes the clouds of the three spin components to deform differently while being spatially separated. In our experiment, we observed that the $m_F=1 (-1)$ component was slightly elongated (compressed) along the $x'$ axis.

To mitigate the influence of the deformation effect on our examination of the spin texture, we introduce an image reconstruction method based on a model that accounts for the stretching and compression caused by the field gradient. In the model, we posit that the magnetic force acting on the $m_F=\pm1$ components is directed radially from a distant zero-field center. Consequently, the $m_F=-1$ component, which seeks low fields, contracts as it is drawn toward the center, while the $m_F=1$ component expands as it moves away. Assuming a constant field gradient magnitude, the classical equation of motion indicates that the $m_F=\pm1$ component initially positioned at $\mathbf{r'}_{\pm1}$ is moved to $\mathbf{r'}(\mathbf{r}_{\pm1}';\mathbf{r_c'},l) = \mathbf{r}_{\pm1}' \pm \frac{\mathbf{r}_{\pm1}'-\mathbf{r_c'}}{|\mathbf{r}_{\pm1}'-\mathbf{r_c'}|}l$, where $\mathbf{r_c'}$ is the position of the zero-field center and $l=|\mathbf{r}'-\mathbf{r}_{\pm1}'|$ is the travel distance during spin separation. Using the relationship between $\mathbf{r'}_{\pm1}$ and $\mathbf{r'}$, the initial density distributions of the $m_F=\pm1$ components can be reconstructed from the measured OD image.

\begin{figure*}[t]
    \includegraphics[width=17cm]{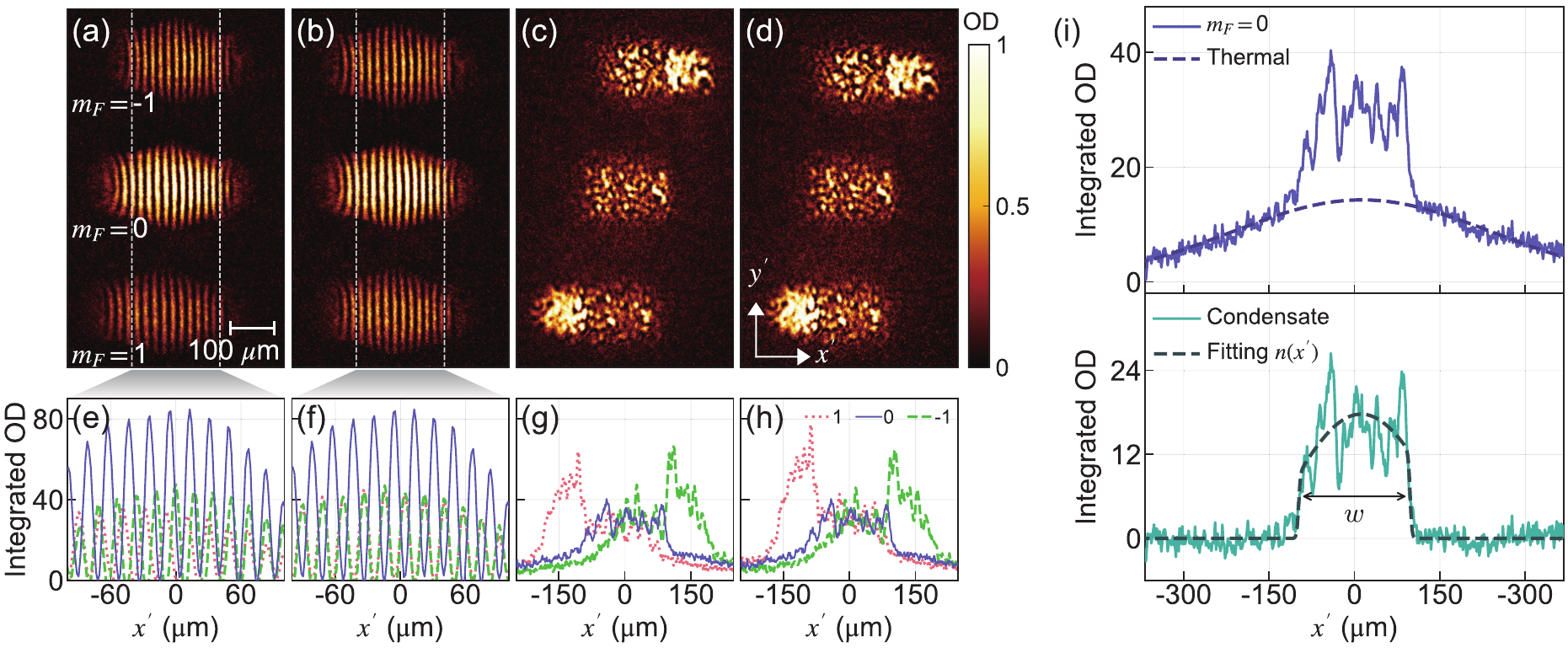}
    \centering
\caption{Stern-Gerlach image reconstruction. (a) Original experimental image of a BEC with helical spin texture, taken after spin separation, and (b) its reconstructed image. (c) Experimental image of a turbulent BEC and (d) its reconstructed image (same image in Fig.~\ref{FIG3}(d)). (e)-(h) Integrated 1D density profiles of the three spin components in (a)-(d).
(i) Determination of the width $w$ of the spin turbulent region from the density profile of the $m_F=$0 spin component. The density profile of the condensate was obtained by removing the thermal contribution using a Gaussian function fitted to the outer region of the entire profile (dashed line in the upper panel). The width $w$ was determined as the distance between the boundary positions $L_1$ and $L_2$ by fitting the condensate profile with an empirical function $n(x')=A\left[\max\left(1-\frac{(x'-x_c')^2}{R_{x'}^2},0\right)\right]^2\left(1+e^{\frac{L_1-x'}{\sigma}}\right)^{-1}\left(1+e^{\frac{x'-L_2}{\sigma}}\right)^{-1}$, where $R_{x'}=210~\mu$m and $\sigma=1.6~\mu$m.}
    \label{FIG.S4}
\end{figure*}

We determine the model parameters $\mathbf{r_c'}$ and $l$, using an image of a BEC with a helical spin texture as a reference, where each spin component shows density stripes, as shown in Fig.~\ref{FIG.S4}(a). The BEC was initially prepared in the $|F= 1,m_F = 0\rangle$ state and its spin state was rotated to $\frac{1}{\sqrt{2}}(|m_F = +1\rangle+|m_F = -1\rangle)$ by applying a $\frac{\pi}{2}$ rf pulse. Subsequently, after an time interval of $\tau=100$~ms under a field gradient of $B'=0.028$~ G/cm along $\hat{\bm{x}}'$, another $\frac{\pi}{2}$ rf pulse was applied to the BEC. Assuming a Thomas-Fermi (TF) density profile for the BEC trapped in a harmonic potential, the column density profiles of the three spin components are expected to be  
\begin{eqnarray}
    n_{m_F=0} (\mathbf{r'}) &=& n_0 \cos^2{(k_{x'} x' +\theta_0 )  \left[\max\left(1- \frac{x'^2}{R_{x'}^2}- \frac{y'^2}{R_{y'}^2},0\right)\right]^{3/2}}, \nonumber\\
    n_{m_F=\pm1} (\mathbf{r'}) &=& \frac{n_0}{2} \sin^2{(k_{x'} x' +\theta_0 ) \left[\max\left(1- \frac{x'^2}{R_{x'}^2}- \frac{y'^2}{R_{y'}^2},0\right)\right]^{3/2}  },
    \label{Ramsey}
\end{eqnarray}
including the Ramsey interference pattern, where $n_0$ is the peak column density of the BEC, $R_{x'(y')}$ is the TF radus along the $x'(y')$ direction, $k_{x'}=\frac{\mu_B B' \tau}{2\hbar}$ is the wave number of the fringe pattern, and $\theta_0$ is the fringe phase at the center of the BEC. We obtain the values of $\mathbf{r_c'}$ and $l$ by fitting the following function to the reference OD image:
\begin{eqnarray}    n(\mathbf{r'};n_0,R_{x'},R_{y'},k_{x'},\theta_0,\mathbf{r_c'},l) = n_{m_F=0}(\mathbf{r'})+\sum_{i=\pm1}\chi_i n_{m_F=i}(\mathbf{r}_i'(\mathbf{r'};\mathbf{r_c'},l)).
\end{eqnarray}
Here $\chi_{\pm1}$ is the OD conversion factor due to changes in cloud size as well as absorption coefficient.

Figure \ref{FIG.S4}(b) shows the density distributions of the three spin components reconstructed from the reference image of Fig.~\ref{FIG.S4}(a). As shown in the integrated 1D density profiles in Figs.~\ref{FIG.S4}(e) and \ref{FIG.S4}(f), the relative shift and deformation between different spin components are well compensated for in the reconstructed image. The same image reconstruction method is applied to the experimental image data taken after the SG spin separation [see Figs.~\ref{FIG.S4}(c) and \ref{FIG.S4}(d)]

\subsection{F.\quad Background field noise contribution}

Fluctuations in the ambient magnetic field can modify the effect of spin driving.
If there is a background field oscillating at the modulation frequency $\nu$, e.g., $\delta B_z(t)=B_\nu \sin (2\pi\nu t +\phi_\nu)$, it interferes with the controlled modulation field, and the driving frequency detuning $\delta$ is modified accordingly as $\delta(t)=\delta_0+ D_{\text{total}} \sin (2\pi\nu t+\phi_{\text{total}})$ with 
\begin{eqnarray}
    &D_{\text{total}}&= \sqrt{D^2+ D_\nu^2 + 2D D_\nu \cos (\phi-\phi_\nu) }, \nonumber \\
    &\tan \phi_{\text{total}} &= \frac{D \sin \phi+D_\nu \sin \phi_\nu}{D \cos \phi +D_\nu \cos \phi_\nu},
\end{eqnarray}
where $D_\nu=\frac{\mu_B}{2\hbar} B_\nu$. Taking into account the contribution of background field fluctuations with frequencies other than $\nu$, we model the threshold detuning as
\begin{eqnarray}
    \delta_\text{th}=\sqrt{\beta_\nu^2 \big(D^2+ 2D D_\nu \cos (\phi-\phi_\nu)\big)+\delta_{\text{th},0}^2},
\end{eqnarray}
where $\delta_{\text{th},0}$ denotes the threshold detuning at $D=0$, purely originating from ambient, uncontrolled field fluctuations. The proportionality factor $\beta_\nu$ represents the efficiency of field modulation to generate turbulence and can be dependent on $\nu$, as observed in the experiment [Fig.~\ref{FIG4}(b)].


\begin{thebibliography}{99}
\bibitem{footnote1_gridsize} The size of the system is $160 \xi_s \times 160 \xi_s$, covered by a $256 \times 256$ grid of equally spaced points, where $\xi_s~\simeq~2.3$ $\mu$m. See \cite{SM} for details.
\bibitem{footnote3_SGimage} In the imaging, the spin axis $\hat{\bm{z}}$ was adiabatically rotated to the spatial axis $\hat{\bm{z}}'$.
\end{thebibliography}
\end{document}